\documentclass[aps,prd,twocolumn,preprintnumbers,superscriptaddress,showpacs]{revtex4-1}
\usepackage{graphicx}

\usepackage{ulem}
\usepackage{color}
\definecolor{My_red}        {cmyk}{0.00,1.00,1.00,0.20}
\def\red{\color{My_red}}

\newcommand{\bmat}{\left(\begin{array}}
\newcommand{\emat}{\end{array}\right)}
\newcommand{\beq}{\begin{equation}}
\newcommand{\eeq}{\end{equation}}

\def\bwt{\begin{widetext}}
\def\ewt{\end{widetext}}
\def\be{\begin{equation}}
\def\ee{\end{equation}}
\def\bea{\begin{eqnarray}}
\def\eea{\end{eqnarray}}
\def\bean{\begin{eqnarray*}}
\def\eean{\end{eqnarray*}}
\def\bary{\begin{array}}
\def\eary{\end{array}}
\def\bit{\begin{itemize}}
\def\eit{\end{itemize}}

\def\su5u1{SU(5) \times U(1)}
\def\fsu5u1{SU(5) \times U(1)'}
\def\so10{SO(10)}
\def\sq20{SO(10) \times SO(10)}

\def\bwt{\begin{widetext}}
\def\ewt{\end{widetext}}
\def\be{\begin{equation}}
\def\ee{\end{equation}}
\def\bea{\begin{eqnarray}}
\def\eea{\end{eqnarray}}
\def\bean{\begin{eqnarray*}}
\def\eean{\end{eqnarray*}}
\def\bary{\begin{array}}
\def\eary{\end{array}}
\def\bit{\begin{itemize}}
\def\eit{\end{itemize}}

\def\su5u1{SU(5) \times U(1)}
\def\fsu5u1{SU(5) \times U(1)'}
\def\so10{SO(10)}
\def\sq20{SO(10) \times SO(10)}

\usepackage[centertags]{amsmath}
\usepackage{amssymb}

\begin{document}

\title{The Return of the King: No-Scale ${\cal F}$-$SU(5)$}

\author{Tianjun Li}

\affiliation{Key Laboratory of Theoretical Physics and Kavli Institute for Theoretical Physics China (KITPC), 
Institute of Theoretical Physics, Chinese Academy of Sciences, Beijing 100190, P. R. China}

\affiliation{ School of Physical Sciences, University of Chinese Academy of Sciences,
  Beijing 100049, P. R. China}

\affiliation{School of Physical Electronics, University of Electronic Science and Technology of China, 
Chengdu 610054, P. R. China}

\author{James A. Maxin}

\affiliation{Department of Physics and Engineering Physics, The University of Tulsa, Tulsa, OK 74104, USA}

\author{Dimitri V. Nanopoulos}

\affiliation{George P. and Cynthia W. Mitchell Institute for Fundamental Physics and Astronomy, Texas A$\&$M University, College Station, TX 77843, USA}

\affiliation{Astroparticle Physics Group, Houston Advanced Research Center (HARC), Mitchell Campus, Woodlands, TX 77381, USA}

\affiliation{Academy of Athens, Division of Natural Sciences, 28 Panepistimiou Avenue, Athens 10679, Greece}

\date{\today}

\begin{abstract}

We revisit the viable parameter space in No-Scale ${\cal F}$-$SU(5)$, examining the Grand Unified Theory within the context of the prevailing gluino mass limits established by the LHC. The satisfaction of both the No-Scale boundary condition and the experimentally measured Standard Model (SM) like Higgs boson mass requires a lower limit on the gluino mass in the model space of about 1.9 TeV, which maybe not coincidentally is the current LHC supersymmetry search bound. This offers a plausible explanation as to why a supersymmetry signal has thus far not been observed at the LHC. On the contrary, since the vector-like flippon particles are relatively heavy due to the strict condition that the supersymmetry breaking
soft term $B_{\mu}$ must vanish at the unification scale, we also cannot address the recently vanished 750~GeV diphoton resonance at the 13~TeV LHC. Therefore, No-Scale ${\cal F}$-$SU(5)$ returns as a King after the spurious 750~GeV diphoton excess was gone with the wind.

\end{abstract}

\pacs{11.10.Kk, 11.25.Mj, 11.25.-w, 12.60.Jv}

\preprint{ACT-06-16,  MI-TH-1627}

\maketitle

\section{Introduction}

Supersymmetry (SUSY) is well acknowledged for the fact it provides a natural solution to the gauge hierarchy problem in the Standard Model (SM). For supersymmetric SMs (SSMs) with $R$-parity in particular, gauge coupling unification can be achieved, the Lightest Supersymmetric Particle (LSP) neutralino serves as a viable dark matter (DM) candidate, and electroweak (EW) gauge symmetry can be broken radiatively due to the large top quark Yukawa coupling, etc. Furthermore, gauge coupling unification strongly implies Grand Unified Theories (GUTs), and SUSY GUTs can be elegantly constructed from superstring theory. As a result, supersymmetry is not only the most promising new physics beyond the SM, but also builds a bridge between the low energy phenomenology and high-energy fundamental physics.

The great success to date at the LHC has been the discovery of a SM-like Higgs boson with an empirically measured mass of $m_h=125.09\pm 0.24$~GeV~\cite{ATLAS, CMS}. Nonetheless, in the Minimal SSM (MSSM), obtaining such a Higgs boson mass requires multi-TeV top squarks with small mixing or TeV-scale top squarks with large mixing~\cite{Carena:2011aa}. However, strong constraints presently exist on the parameter space in the SSMs from LHC SUSY searches. For instance, the most recent search bounds on the gluino (${\tilde g}$) mass show that it is heavier than about 1.9 TeV, whereas the light stop (${\tilde t}_1$) mass is heavier than about 900 GeV~\cite{WAdam-ICHEP}. Therefore, naturalness in the SSMs is challenged from both the Higgs boson mass and the LHC SUSY searches. On the other hand, the ATLAS~\cite{bib:ATLAS_diphoton} and CMS~\cite{bib:CMS_diphoton} Collaborations announced in December 2015 an excess of events in the diphoton channel with invariant mass of about 750~GeV at the 13~TeV LHC run II, though this dubious excess was proven to be only a statistical fluctuation in recent LHC data~\cite{Khachatryan:2016yec}. Hence, any natural candidate for the GUT model of our universe must also be consistent with the vanishing of this diphoton resonance.

To achieve the string-scale gauge coupling unification, we proposed the testable flipped $SU(5)\times U(1)_X$ models~\cite{smbarr, dimitri, AEHN-0} with TeV-scale vector-like particles~\cite{Jiang:2006hf}, dubbed flippons. Subsequently, we constructed these flipped $SU(5)$ models from local F-theory model building~\cite{Jiang:2008yf, Jiang:2009za}, where these models can be obtained in free-fermionic string constructions as well~\cite{LNY}. The models were thus referred to as ${\cal F}$-$SU(5)$. 
A brief review of the ``miracles''~\cite{Li:2011ab} of flippons in ${\cal F}$-$SU(5)$ is now in order. First, the lightest CP-even Higgs boson mass can be lifted to 125~GeV easily because of the one-loop contributions from the Yukawa couplings between the flippons and Higgs fields~\cite{Huo:2011zt, Li:2011ab}. In the present work, this will only be relevant for those lighter regions of the model space which have already been excluded by the LHC, hence, we shall assume the minimal Yukawa couplings amongst the flippons and Higgs fields. Second, although the dimension-five proton decays mediated by colored Higgsinos are highly suppressed due to the missing partner mechanism and TeV-scale $\mu$ term, the dimension-six proton decays via the heavy gauge boson exchanges are within the reach of the future proton decay experiments such as the Hyper-Kamiokande experiment. The key point is that the $SU(3)_C\times SU(2)_L$ gauge couplings are still unified at the traditional GUT scale while the unified gauge couplings become larger due to vector-like particle contributions~\cite{Li:2009fq, Li:2010dp}. Therefore, the ${\cal F}$-$SU(5)$ models differ from the minimal flipped $SU(5)\times U(1)_X$ model, whose proton lifetime is too lengthy 
for the future proton decay experiments. Third, we can consider No-Scale supergravity~\cite{Cremmer:1983bf} as a result of the string model building. More specifically, the lightest neutralino fulfills the role of the LSP and is lighter than the light stau due to the longer running of the Renormalization Group Equations (RGEs), providing the LSP neutralino as a dark matter candidate~\cite{Li:2010ws, Li:2010mi, Li:2011xua}. Fourth, given No-Scale supergravity, there exists a distinctive mass ordering $M({\tilde t}_1) < M({\tilde g}) < M({\tilde q}) $ of a light stop and gluino in No-Scale ${\cal F}$-$SU(5)$, with both substantially lighter than all other squarks (${\tilde q}$)~\cite{Li:2010ws, Li:2010mi, Li:2011xua}. A primary consequence of this SUSY spectrum mass pattern at the LHC is the prediction of large multijets events~\cite{Li:2011hya}. Fifth, with a merging of both No-Scale supergravity and the Giudice-Masiero (GM) mechanism~\cite{Giudice:1988yz}, the supersymmetry electroweak fine-tuning problem can be elegantly solved rather naturally~\cite{Leggett:2014mza, Leggett:2014hha}. Conversely, to satisfy the No-Scale boundary condition $B_{\mu} = 0$ and obtain the experimentally observed SM like Higgs boson mass, we find that the flippons are required to be relatively heavy, and as such we cannot explain the recently vanished 750~GeV diphoton resonance at the 13~TeV LHC, which seemed to prefer rather light vector-like particle masses. In conclusion, No-Scale ${\cal F}$-$SU(5)$ returns post disappearance of the 750~GeV diphoton excess. In this paper, we revisit and update the viable parameter space of No-Scale ${\cal F}$-$SU(5)$, exhibiting that consistency with both No-Scale boundary conditions and the experimentally measured SM like Higgs boson mass necessitates a lower bound on the gluino mass in the model space of around 1.9 TeV, which perhaps not coincidentally is the current LHC supersymmetry search bound, presenting a plausible explanation for the absence to date of a definitive SUSY signal at the LHC.

\section{Brief Review of No-Scale ${\cal F}$-$SU(5)$ Models }

We now briefly review the minimal flipped $SU(5)$ model~\cite{smbarr, dimitri, AEHN-0}. The gauge group for the flipped $SU(5)$ model is $SU(5)\times U(1)_{X}$, which can be embedded into the $SO(10)$ model. We define the generator $U(1)_{Y'}$ in $SU(5)$ as 
\bea 
T_{\rm U(1)_{Y'}}={\rm diag} \left(-\frac{1}{3}, -\frac{1}{3}, -\frac{1}{3},
 \frac{1}{2},  \frac{1}{2} \right).
\label{u1yp}
\eea
and the hypercharge is given by
\bea
Q_{Y} = \frac{1}{5} \left( Q_{X}-Q_{Y'} \right).
\label{ycharge}
\eea
There are three families of the SM fermions whose quantum numbers under $SU(5)\times U(1)_{X}$ are respectively
\bea
F_i={\mathbf{(10, 1)}},~ {\bar f}_i={\mathbf{(\bar 5, -3)}},~
{\bar l}_i={\mathbf{(1, 5)}},
\label{smfermions}
\eea
where $i=1, 2, 3$. The SM particle assignments in $F_i$, ${\bar f}_i$ and ${\bar l}_i$ are
\bea
F_i=(Q_i, D^c_i, N^c_i),~{\overline f}_i=(U^c_i, L_i),~{\overline l}_i=E^c_i~,~
\label{smparticles}
\eea
where $Q_i$ and $L_i$ are respectively the superfields of the left-handed quark and lepton doublets, $U^c_i$, $D^c_i$, $E^c_i$ and $N^c_i$ are the $CP$ conjugated superfields for the right-handed up-type quarks, down-type quarks, leptons and neutrinos, respectively. To generate the heavy right-handed neutrino masses, we can introduce three SM singlets $\phi_i$.

The breaking of the GUT and electroweak gauge symmetries results from introduction of two pairs of Higgs representations
\begin{eqnarray}
H&=&{\mathbf{(10, 1)}},~{\overline{H}}={\mathbf{({\overline{10}}, -1)}}, \nonumber \\
h&=&{\mathbf{(5, -2)}},~{\overline h}={\mathbf{({\bar {5}}, 2)}}.
\label{Higgse1}
\end{eqnarray}
We label the states in the $H$ multiplet by the same symbols as in the $F$ multiplet, and for ${\overline H}$ we just add ``bar'' above the fields. Explicitly, the Higgs particles are
\bea
H=(Q_H, D_H^c, N_H^c)~,~
{\overline{H}}= ({\overline{Q}}_{\overline{H}}, {\overline{D}}^c_{\overline{H}}, 
{\overline {N}}^c_{\overline H})~,~\,
\label{Higgse2}
\eea
\bea
h=(D_h, D_h, D_h, H_d)~,~
{\overline h}=({\overline {D}}_{\overline h}, {\overline {D}}_{\overline h},
{\overline {D}}_{\overline h}, H_u)~,~\,
\label{Higgse3}
\eea
where $H_d$ and $H_u$ are one pair of Higgs doublets in the MSSM. We also add one SM singlet $\Phi$.

The $SU(5)\times U(1)_{X}$ gauge symmetry is broken down to the SM gauge symmetry by introduction of the following Higgs superpotential at the GUT scale
\bea
{\it W}_{\rm GUT}=\lambda_1 H H h + \lambda_2 {\overline H} {\overline H} {\overline
h} + \Phi ({\overline H} H-M_{\rm H}^2)~.~ 
\label{spgut} 
\eea
There is only one F-flat and D-flat direction, which can always be rotated along the $N^c_H$ and ${\overline {N}}^c_{\overline H}$ directions. Therefore, we obtain $<N^c_H>=<{\overline {N}}^c_{\overline H}>=M_{\rm H}$. In addition, the superfields $H$ and ${\overline H}$ are eaten and acquire large masses via the supersymmetric Higgs mechanism, except for $D_H^c$ and ${\overline {D}}^c_{\overline H}$. Furthermore, the superpotential terms $ \lambda_1 H H h$ and $ \lambda_2 {\overline H} {\overline H} {\overline h}$ couple the $D_H^c$ and ${\overline {D}}^c_{\overline H}$ with the $D_h$ and ${\overline {D}}_{\overline h}$, respectively, to form the massive eigenstates with masses $2 \lambda_1 <N_H^c>$ and $2 \lambda_2 <{\overline {N}}^c_{\overline H}>$. As a consequence, we naturally have the doublet-triplet splitting due to the missing partner mechanism~\cite{AEHN-0}. The triplets in $h$ and ${\overline h}$ only have small mixing through the $\mu$ term, hence, the Higgsino-exchange mediated proton decay is negligible, {\it i.e.}, there is no dimension-5 proton decay problem. 

String-scale gauge coupling unification~\cite{Jiang:2006hf, Jiang:2008yf, Jiang:2009za} is achieved by the introduction of the following vector-like particles (flippons) at the TeV scale
\begin{eqnarray}
&& XF ={\mathbf{(10, 1)}}~,~{\overline{XF}}={\mathbf{({\overline{10}}, -1)}}~,~\\
&& Xl={\mathbf{(1, -5)}}~,~{\overline{Xl}}={\mathbf{(1, 5)}}~.~\,
\end{eqnarray}
The particle content from the decompositions of $XF$, ${\overline{XF}}$, $Xl$, and ${\overline{Xl}}$ under the SM gauge symmetry are
\begin{eqnarray}
&& XF = (XQ, XD^c, XN^c)~,~ {\overline{XF}}=(XQ^c, XD, XN)~,~\\
&& Xl= XE~,~ {\overline{Xl}}= XE^c~.~
\end{eqnarray}
Under the $SU(3)_C \times SU(2)_L \times U(1)_Y$ gauge symmetry, the quantum numbers for the extra vector-like particles are
\begin{eqnarray}
&& XQ={\mathbf{(3, 2, \frac{1}{6})}}~,~
XQ^c={\mathbf{({\bar 3}, 2,-\frac{1}{6})}} ~,~\\
&& XD={\mathbf{({3},1, -\frac{1}{3})}}~,~
XD^c={\mathbf{({\bar 3},  1, \frac{1}{3})}}~,~\\
&& XN={\mathbf{({1},  1, {0})}}~,~
XN^c={\mathbf{({1},  1, {0})}} ~,~\\
&& XE={\mathbf{({1},  1, {-1})}}~,~
XE^c={\mathbf{({1},  1, {1})}}~.~\,
\label{qnum}
\end{eqnarray}

Mass degeneracy of the superpartners has not been observed, so SUSY must be broken around the TeV scale. In GUTs with gravity mediated supersymmetry breaking, called the supergravity models, we can fully characterize the supersymmetry breaking soft terms by four universal parameters (gaugino mass $M_{1/2}$, scalar mass $M_0$, trilinear soft term $A$, and the low energy ratio of Higgs vacuum expectation values (VEVs) $\tan\beta$), plus the sign of the Higgs bilinear mass term $\mu$.

No-Scale Supergravity was proposed~\cite{Cremmer:1983bf} to solve the cosmological flatness problem, as the subset of supergravity models which satisfy the following three constraints: i) the vacuum energy vanishes automatically due to the suitable
 K\"ahler potential; ii) at the minimum of the scalar potential there exist flat directions that leave the gravitino mass $M_{3/2}$ undetermined; iii) the quantity ${\rm Str} {\cal M}^2$ is zero at the minimum. If the third condition were not true, large one-loop corrections would force $M_{3/2}$ to be either identically zero or of the Planck scale. A simple K\"ahler potential that satisfies the first two conditions is~\cite{Cremmer:1983bf}
\begin{eqnarray} 
K &=& -3 {\rm ln}( T+\overline{T}-\sum_i \overline{\Phi}_i
\Phi_i)~,~
\label{NS-Kahler}
\end{eqnarray}
where $T$ is a modulus field and $\Phi_i$ are matter fields, which parameterize the non-compact $SU(N,1)/SU(N) \times U(1)$ coset space. The third condition is model dependent and can always be satisfied in principle~\cite{Ferrara:1994kg}. For the simple K\"ahler potential in Eq.~(\ref{NS-Kahler}) we automatically obtain the No-Scale boundary condition $M_0=A=B_{\mu}=0$ at the ultimate unification scale $M_{\cal F}$, while the sole model parameter $M_{1/2}$ is allowed, and indeed required for SUSY breaking. Because the minimum of the electroweak (EW) Higgs potential $(V_{EW})_{min}$ depends on $M_{3/2}$,  the gravitino mass is determined by the equation $d(V_{EW})_{min}/dM_{3/2}=0$. Thus, the supersymmetry breaking scale is determined dynamically. No-Scale supergravity can be realized in the compactification of the weakly coupled heterotic string theory~\cite{Witten:1985xb} and the compactification of M-theory on $S^1/Z_2$ at the leading order~\cite{Li:1997sk}.

Given that the $B_{\mu}$ parameter is determined at the $M_{\cal F}$ scale from the No-Scale boundary conditions, this in principle determines tan$\beta$, though in the analytical procedure to follow here we use a consistency check to uncover those values of tan$\beta$ that are consistent with $B_{\mu}(M_{\cal F})$ = 0, rather than solve for the explicit values of tan$\beta$ directly. The scale at which the vector-like flippon particles decouple is defined as $M_V$, and as we shall show, is a function of $M_{1/2}$ via the RGE running. So in effect, all parameters are reduced to a dependence on $M_{1/2}$, providing a genuine one-parameter model.

\section{Numerical Results}

\begin{figure}[htp]
        \centering
        \includegraphics[width=0.5\textwidth]{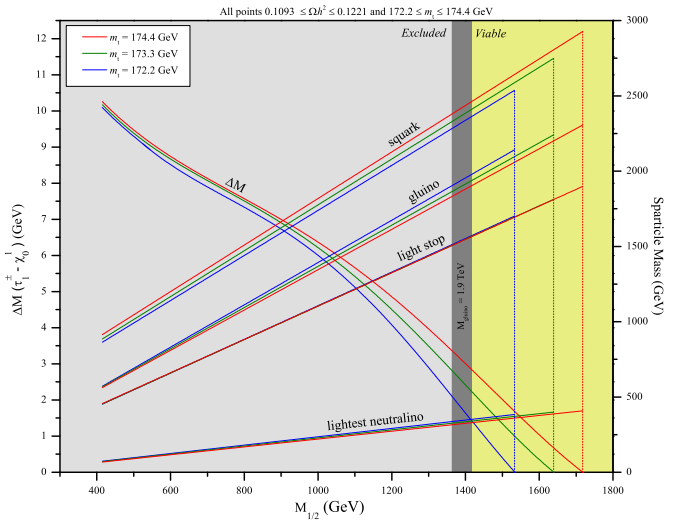}
        \caption{Depiction of the SUSY spectrum masses for the lightest neutralino $\widetilde{\chi}_1^0$, light stop $\widetilde{t}_1$, gluino $\widetilde{g}$, right-handed up squark $\widetilde{u}_R$, and mass difference $\Delta M = M(\widetilde{\tau}_1^{\pm}) - M(\widetilde{\chi}_1^0)$ as a function of the sole model parameter $M_{1/2}$ for three discrete values of the top quark mass $m_t = \{172.2,~173.3,~174.4\}$~GeV. All points included adhere to the constraints on the relic density $0.1093 \le \Omega h^2 \le 0.1221$ and top quark mass $172.2 \le m_t \le 174.4$~GeV. Contours shown are numerical fits, though the full compilation of points only show a small tolerance around these fitted lines due to the strict condition $| B_{\mu} | \le$ 1~GeV, thus the numerical fits are rather representative of the actual model space. Therefore, the relationship between the SUSY masses and $M_{1/2}$ is indeed a linear function as illustrated. The plot space is segregated into viable and excluded as established by the LHC given the current gluino mass limit of about 1.9~TeV.}
        \label{fig:sparticles}
\end{figure}

\begin{figure}[htp]
        \centering
        \includegraphics[width=0.5\textwidth]{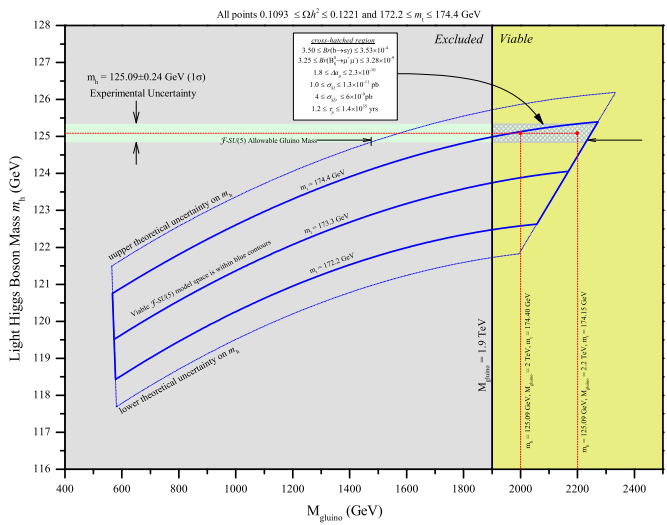}
        \caption{Illustration of the gluino mass $M_{\widetilde{g}}$ as a function of the lightest Higgs boson mass $m_{h}$. All points included adhere to the constraints on the relic density $0.1093 \le \Omega h^2 \le 0.1221$ and top quark mass $172.2 \le m_t \le 174.4$~GeV. Contours shown are for three discrete values of the top quark mass $m_t = \{172.2,~173.3,~174.4\}$~GeV. Also displayed is the 1.5~GeV theoretical uncertainty on the calculations of the light Higgs boson mass and the 1$\sigma$ experimental uncertainty on the light Higgs mass of $m_h = 125.09 \pm 0.24$~GeV. The union of the 1$\sigma$ experimental uncertainty with the theoretical uncertainty on our calculations generates a viable gluino mass range in the model of $1.5 \lesssim M_{\widetilde{g}} \lesssim 2.3$~TeV, though the intersection of the central experimental and theoretical values provides a rather compelling fundamental link between the gluino and light Higgs boson masses, as evidenced by the cross-hatched region and the two points highlighted therein. Further illuminated are the rare-decay processes, direct dark matter detection cross-sections, and proton decay rates computed for the cross-hatched region. The plot space is segregated into viable and excluded as established by the LHC given the current gluino mass limit of about 1.9~TeV. A minimal coupling of the vector-like flippon multiplets to the light Higgs boson is assumed.}
        \label{fig:higgs_gluino}
\end{figure}

\begin{table*}[htp]
\centering
\footnotesize
\caption{Sample No-Scale ${\cal F}$-$SU(5)$ benchmark points for $m_t$ = 173.3~GeV and $m_t$ = 174.4~GeV that satisfy all experimental constraints imposed by the LHC and other essential experiments. All masses are in GeV. The numerical values given for $\Delta a_{\mu}$ are $\times 10^{-10}$, $Br(b \rightarrow s \gamma)$ are $\times 10^{-4}$,  $Br(B_s^0 \rightarrow \mu^+ \mu^-)$ are $\times 10^{-9}$, spin-independent cross-sections $\sigma_{SI}$ are $\times 10^{-11}$~pb, spin-dependent cross-sections $\sigma_{SD}$ are $\times 10^{-9}$~pb, and proton decay rate $p \to e^+ \pi^0$ are in units of $10^{35}$ years. The $\Delta M$ represents the mass difference between the light stau and lightest neutralino, given here to sufficient precision.}
\begin{tabular}{|c|c|c|c||c|c|c|c|c|c|c||c|c|c|c|c|c|c|} \hline
$M_{1/2}$&$M_V$&${\rm tan}\beta$&$m_{\rm top}$&$M_{\chi_1^0}$&$M_{\widetilde{\tau}^{\pm}}$&$\Delta M$&$M_{\widetilde{t}_1}$&$M_{\widetilde{u}_R}$&$	{\bf \red M_{\widetilde{g}}}$&$	{\bf \red M_h}$&$\Omega h^2$&$\Delta a_{\mu}$&$Br(b \rightarrow s \gamma)$&$Br(B_s^0 \rightarrow \mu^+ \mu^-)$&$\sigma_{SI}$&$\sigma_{SD}$&$\tau_p$\\ \hline \hline
$	1532	$&$	80861	$&$	24.95	$&$	173.3	$&$	371	$&$	372	$&$	1.10	$&$	1693	$&$	2585	$&$	2095$&$	124.00	$&$	0.1177	$&$	2.24	$&$	3.50	$&$	3.21	$&$	1.5	$&$	6.2	$&$	1.34	$	\\ \hline
$	1577	$&$	92472	$&$	25.03	$&$	173.3	$&$	385	$&$	385	$&$	0.73	$&$	1742	$&$	2648	$&$	2158$&$	124.05	$&$	0.1184	$&$	2.13	$&$	3.51	$&$	3.21	$&$	1.4	$&$	5.7	$&$	1.37	$	\\ \hline
$	1592	$&$	96597	$&$	25.05	$&$	173.3	$&$	389	$&$	390	$&$	0.59	$&$	1758	$&$	2669	$&$	2179$&$	124.06	$&$	0.1182	$&$	2.10	$&$	3.51	$&$	3.20	$&$	1.4	$&$	5.6	$&$	1.38	$	\\ \hline \hline
$	1514	$&$	30195	$&$	24.69	$&$	174.4	$&$	353	$&$	355	$&$	1.77	$&$	1675	$&$	2619	$&$	{\bf \red 2031}$&$	{\bf \red 125.17}	$&$	0.1190	$&$	2.18	$&$	3.51	$&$	3.27	$&$	1.2	$&$	5.1	$&$	1.23	$	\\ \hline
$	1569	$&$	35122	$&$	24.79	$&$	174.4	$&$	369	$&$	370	$&$	1.20	$&$	1734	$&$	2699	$&$	{\bf \red 2107}$&$	{\bf \red 125.28}	$&$	0.1180	$&$	2.06	$&$	3.52	$&$	3.26	$&$	1.1	$&$	4.7	$&$	1.26	$	\\ \hline
$	1620	$&$	40350	$&$	24.88	$&$	174.4	$&$	383	$&$	384	$&$	0.79	$&$	1787	$&$	2771	$&$	{\bf \red 2177}$&$	{\bf \red 125.37}	$&$	0.1186	$&$	1.95	$&$	3.52	$&$	3.26	$&$	1.0	$&$	4.3	$&$	1.30	$	\\ \hline
$	1653	$&$	44180	$&$	24.94	$&$	174.4	$&$	393	$&$	394	$&$	0.50	$&$	1822	$&$	2817	$&$	{\bf \red 2222}$&$	{\bf \red 125.39}	$&$	0.1183	$&$	1.89	$&$	3.53	$&$	3.25	$&$	1.0	$&$	4.1	$&$	1.32	$	\\ \hline
$	1685	$&$	48307	$&$	25.00	$&$	174.4	$&$	403	$&$	403	$&$	0.21	$&$	1857	$&$	2863	$&$	{\bf \red 2268}$&$	{\bf \red 125.45}	$&$	0.1172	$&$	1.83	$&$	3.53	$&$	3.25	$&$	1.0	$&$	3.9	$&$	1.34	$	\\ \hline
\end{tabular}
\label{tab:OPM1}
\end{table*}

The LHC will soon increase its reach to probe for a 2 TeV gluino and beyond, so we update and compute the precise upper boundary of the No-Scale ${\cal F}$-$SU(5)$ parameter space, extending the analysis of Ref.~\cite{Li:2013naa}. This upper limit is entirely defined by the requirement of neutralino dark matter. Our first constraints imposed are the WMAP 9-year~\cite{Hinshaw:2012aka} and 2015 Planck~\cite{Ade:2015xua} 1$\sigma$ relic density measurements, where we constrain the model to be consistent with both data sets, imposing limits of $0.1093 \le \Omega h^2 \le 0.1221$, as well as a sufficient range of the top quark mass around the world average~\cite{CDF:2013jga}, implementing limits in our analysis of $172.2 \le m_t \le 174.4$~GeV. These requirements on dark matter abundance and the top quark mass establish a hard upper boundary on the model space, as shown in FIG.~\ref{fig:sparticles}. The plot space in FIG.~\ref{fig:sparticles} and all subsequent figures in this work are segregated into those two regions separated by the present exclusion boundary established by the LHC of $M_{\widetilde{g}} \gtrsim 1.9$~TeV. The lines in FIG.~\ref{fig:sparticles} represent a numerical fit to the viable points in the model space for three discrete values of the top quark mass $m_t = \{172.2,~173.3,~174.4\}$~GeV after imposing the noted WMAP9, Planck, and top mass constraints, in addition to the strict vanishing of the $B_{\mu}$ parameter at the $M_{\cal F}$ scale, applied as $| B_{\mu} | \le$ 1~GeV, which is consistent with the induced variation from fluctuation of the strong coupling within its error bounds, and likewise with the expected scale of radiative EW corrections. While there is a rather small tolerance around these fitted lines resulting from the very narrow condition $| B_{\mu} | \le$ 1~GeV, the actual points themselves for the sparticle masses are definitively linear as shown in the figure. It is clear that the mass difference between the lightest neutralino and light stau, defined here as $\Delta M = M(\widetilde{\tau}_1^{\pm}) - M(\widetilde{\chi}_1^0)$, approaches zero and subsequently further decreases to negative values. The requirement of neutralino dark matter necessitates $\Delta M \ge 0$, therefore providing a maximum gluino mass of 2.27 TeV, given an explicit WMAP9 and Planck 1$\sigma$ relic density constraint on the model space. If the relic density measurements are relaxed, then the upper boundary of the model space could be extended. However, for the purposes of this work, we shall strictly adhere to the 1$\sigma$ ranges on the WMAP9 and 2015 Planck measurements. The FIG.~\ref{fig:sparticles} also exhibits the rather elegant proportionality of the entire SUSY spectrum as a function of the sole model parameter, $M_{1/2}$. This is illustrated in FIG.~\ref{fig:sparticles} for the lightest neutralino $\widetilde{\chi}_1^0$, light stop $\widetilde{t}_1$, gluino $\widetilde{g}$, and right-handed up squark $\widetilde{u}_R$, where all are linear functions of the sole model parameter $M_{1/2}$.
The naive puzzle is that $\Delta M$ may be linearly proportional to $M_{1/2}$ as well, which will be addressed 
in the last part of this Section.

\begin{figure}[htp]
        \centering
        \includegraphics[width=0.5\textwidth]{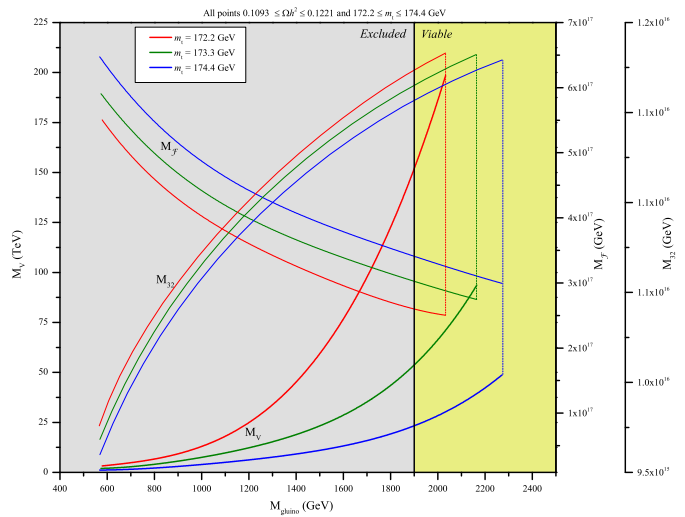}
        \caption{Representation of the three significant mass scales in No-Scale ${\cal F}$-$SU(5)$ as a function of the gluino mass. Included here are the vector-like flippon mass scale $M_V$, the $SU(3)_C\times SU(2)_L$ secondary unification scale $M_{32}$, and the $SU(5) \times U(1)_X$ unification scale $M_{\cal F}$. All points included adhere to the constraints on the relic density $0.1093 \le \Omega h^2 \le 0.1221$ and top quark mass $172.2 \le m_t \le 174.4$~GeV. Contours shown are for three discrete values of the top quark mass $m_t = \{172.2,~173.3,~174.4\}$~GeV. Contours shown are numerical fits, though the full compilation of points only show a small tolerance around these fitted lines due to the strict condition $| B_{\mu} | \le$ 1~GeV, thus the numerical fits are rather representative of the actual model space. The plot space is segregated into viable and excluded as established by the LHC given the current gluino mass limit of about 1.9~TeV.}
        \label{fig:MV}
\end{figure}

The chief thrust of this work though can be found via an examination of FIG.~\ref{fig:higgs_gluino}, prominently displaying the remarkable relationship between the gluino mass and light Higgs boson mass $m_h$ in No-Scale ${\cal F}$-$SU(5)$. In fact, the light Higgs boson mass experiences a smooth increase with increasing gluino mass, with both the gluino and Higgs boson mass entering into their experimentally viable ranges simultaneously. Indeed, the theoretical calculation of the light Higgs boson mass in the model does not reach the 1$\sigma$ experimental range of $m_h = 125.09 \pm 0.24$~GeV~\cite{ATLAS,CMS} until the calculated gluino mass surpasses 1.9~TeV! Hence, given a potential substantiation of ${\cal F}$-$SU(5)$ in the near future at the LHC, it is of no surprise that  definitive signals of SUSY have not been uncovered yet. The reach of the LHC is just now presently entering into the viable model space that computes the correct light Higgs boson mass. We base this analysis on the central value of the experimental Higgs mass of $m_h = 125.09$~GeV, though even the narrow 1$\sigma$ tolerance of $\pm 0.24$~GeV delivers the same message that the LHC is currently probing the viable region of the model space where a SUSY discovery would be expected. It should be noted that our Higgs boson mass calculations assume a minimal coupling of the flippon vector-like multiplets. Although this has no effect on the Higgs mass calculations for a gluino mass greater than 1.9 TeV due to the rather large flippon mass $M_V$ required to satisfy the theoretical constraint of $| B_{\mu} | \le$ 1~GeV, it would though provide a larger contribution to those excluded regions for $M_{\widetilde{g}} \lesssim 1$~TeV, raising the Higgs mass to about 125~GeV for these lighter regions of the model space~\cite{Li:2011ab}. 

The SUSY mass spectra, relic density, rare decay processes, and direct dark matter detection cross-sections are calculated with {\tt MicrOMEGAs~2.1}~\cite{Belanger:2008sj} utilizing a proprietary modification of the {\tt SuSpect~2.34}~\cite{Djouadi:2002ze} codebase to run flippon and No-Scale ${\cal F}$-$SU(5)$ enhanced RGEs. The theoretically computed light Higgs boson mass consists of only the 1-loop and 2-loop SUSY contributions, primarily from the coupling to the light stop. We also take into account a theoretical uncertainty on our calculations of 1.5 GeV, shown for the model space extremes in FIG.~\ref{fig:higgs_gluino}, though for clarity we base our primary conclusions stated here on our centrally computed value. The theoretical uncertainty of 1.5 GeV in our calculations gives a lower bound on the gluino mass in the model space of about 1.5 TeV, and an upper bound just above 2.3 TeV, ironically the range of gluino mass currently under probe at the LHC.

The resiliency of No-Scale ${\cal F}$-$SU(5)$ is exemplified by a persistent consistency with all presently running experiments. The slender cross-hatched region depicted in FIG.~\ref{fig:higgs_gluino} highlights the viable region currently under test and the associated rare decay, direct detection, and proton lifetime numerical results, which satisfy the experimental constraints on the branching ratio of the rare b-quark decay of $Br(b \to s \gamma) = (3.43 \pm 0.21^{stat}~ ±\pm 0.24^{th} \pm 0.07^{sys}) \times 10^{-4}$~\cite{HFAG}, the branching ratio of the rare B-meson decay to a dimuon of $Br(B_s^0 \to \mu^+ \mu^-) = (2.9 \pm 0.7 \pm 0.29^{th}) \times 10^{-9}$~\cite{CMS:2014xfa}, the 3$\sigma$ intervals around the SM value and experimental measurement of the SUSY contribution to the anomalous magnetic moment of the muon of $-17.7 \times10^{-10} \le \Delta a_{\mu} \le 43.8 \times 10^{-10}$~\cite{Aoyama:2012wk}, limits on spin-independent cross-sections for neutralino-nucleus interactions derived by the LUX experiment~\cite{Akerib:2013tjd}, limits on the proton spin-dependent cross-sections by the COUPP Collaboration~\cite{Behnke:2012ys} and XENON100 Collaboration~\cite{Aprile:2013doa}, and current limits of about $1.7 \times 10^{34}$~yrs on the proton decay rate $p \to e^+ \pi^0$ in the context of flipped $SU(5)$ grand unification~\cite{Takhistov:2016eqm}. In short, there is no prominent SUSY related experiment that No-Scale ${\cal F}$-$SU(5)$ is not consistent with. Results of all these detailed calculations along with the primary sparticle masses are listed in TABLE~\ref{tab:OPM1} for a set of eight viable sample benchmark points for a given set of input parameters $(M_{1/2},~M_V,~m_t,~{\rm tan}\beta)$. While a top quark mass of $m_t = 173.3$~GeV does generate a Higgs mass just within its lower experimental 2$\sigma$ boundary of about $m_h \simeq 124$~GeV, certainly the better fit to the 1$\sigma$ Higgs mass experimental value is given by a top quark mass of $m_t \simeq 174.4$~GeV, as highlighted by the two points annotated in FIG.~\ref{fig:higgs_gluino}. The striking correlation between the gluino and Higgs masses in their respective columns in TABLE~\ref{tab:OPM1} is unmistakeable, presenting a rather natural solution to the chronic dilemma at the LHC regarding the absence thus far of a conclusive SUSY signal. 

From TABLE~\ref{tab:OPM1} it can be seen that the mass difference $\Delta M$ between the light stau and lightest neutralino for the viable region we analyze in this work spans from a degenerate light stau and lightest neutralino at the upper bound of the model space, to a mass delta equivalent to the tau mass $\tau^{\pm} = 1.777$~GeV. The branching fraction of a light stau decay to the lightest neutralino $\widetilde{\tau}_1^{\pm} \to \tau^{\pm} + \widetilde{\chi}_1^0$ is 100\%, therefore, in this particular region we study, this decay mode consists of an off-shell tau.

The recently excluded possibility of a 750~GeV diphoton resonance seemed to prefer rather light vector-like masses in order to generate the temporarily observed cross section~\cite{Li:2016xcj,Li:2016tqf}. In the event the diphoton resonance would have been confirmed, this requirement of light vector-like masses would have surely excluded our one-parameter version of No-Scale ${\cal F}$-$SU(5)$ since the viable vector-like mass $M_V$ is larger than about 23 TeV from FIG.~\ref{fig:MV} due to mostly the $B_{\mu}=0$ condition. However, as would be necessary for No-Scale ${\cal F}$-$SU(5)$ to remain viable as a natural GUT candidate, the diphoton resonance curiously faded into oblivion. The reasoning behind the assertion noted above is depicted in FIG.~\ref{fig:MV}, delineating the dependent relationship between the vector-like flippon mass decoupling scale $M_V$, the $SU(3)_C\times SU(2)_L$ secondary unification scale $M_{32}$, and the $SU(5) \times U(1)_X$ unification scale $M_{\cal F}$. This figure graphically illustrates the required largeness of $M_V$ near the upper boundary of the model space when the strict WMAP9, 2015 Planck, and world average top quark mass constraints are applied. The dominant effect leading to such large numerical values of $M_V$ relates to the rather tight theoretical constraint $| B_{\mu} | \le$ 1~GeV, where contours of constant $B_{\mu}$ are generated as a function of the gluino mass, as shown in FIG.~\ref{fig:MV} for the specific constant value of $B_{\mu} \simeq 0$. While the top quark mass $m_t$ and tan$\beta$ induce smaller corrections to these contours that must be taken into account when adhering to the 1$\sigma$ relic density and top mass constraints, the dominant effect certainly resides with the $B_{\mu} \simeq 0$ condition. In fact, those regions below the $M_V$ contours in FIG.~\ref{fig:MV} for smaller $M_V$ produce contours of constant $B_{\mu}$ for $ B_{\mu} > 0$, with values as large as $B_{\mu} \sim 10$, whereas those regions above the $M_V$ contours in FIG.~\ref{fig:MV} for larger $M_V$ produce contours of constant $B_{\mu}$ for $ B_{\mu} < 0$, with values as small as $B_{\mu} \sim -20$. Regarding the non-linear proportionality of $\Delta M$ as a function of $M_{1/2}$,  
it can be seen in FIG.~\ref{fig:MV} that when  $M_{1/2}$ or $M_{\tilde g}$ increases, $M_V$ will also increase and thus $M_{\cal F}$ decreases. With smaller $M_{\cal F}$, the renormalization scale range for RGE running becomes shorter and then $\Delta M$ will decrease as well.

\section{Conclusions}

We revisited the viable parameter space in No-Scale ${\cal F}$-$SU(5)$, examining the GUT model given the updated gluino mass limit of $M_{\widetilde{g}} \gtrsim 1.9$~TeV established by the LHC. To satisfy both the No-Scale boundary condition and the experimentally measured Higgs boson mass, we discovered that the lower limit on the gluino mass in the model space is curiously also about 1.9 TeV, rather similar to the current LHC supersymmetry search bound. This does present a legitimate explanation as to why no supersymmetry signal has been observed at the LHC to date. Moreover, due to the fact that the vector-like flippon particles are relatively heavy, primarily resulting from the No-Scale boundary condition $B_{\mu}=0$ at the unification scale, the model appropriately excludes the recently fizzled 750~GeV diphoton resonance at the 13~TeV LHC, as is required of any viable GUT candidate. The natural union of the LHC gluino mass limit and experimentally measured Higgs boson mass in No-Scale ${\cal F}$-$SU(5)$ serves as a prime region for SUSY probing at the LHC, given also this region's quite favorable consistency with all other essential SUSY experiments involving relic density observations, rare decay processes, direct dark matter detection, and proton lifetime measurements. While SUSY enthusiasts have endured several setbacks over the prior few years amidst the discouraging results at the LHC in the search for supersymmetry, it is axiomatic that as a matter of course, great triumph emerges from momentary defeat. As the precession of null observations at the LHC has surely dampened the spirits of SUSY proponents, the conclusion of our analysis here indicates that the quest for SUSY may just be getting interesting.

%%%%%%%%%%%%%%%%%%%%%%%%%%%%%%%%%%%%%%%%%%%%%%%%%%%

\begin{acknowledgments}

The computing for this project was performed at the Tandy Supercomputing Center, using dedicated resources provided by The University of Tulsa. This research was supported in part by the Natural Science Foundation of China under grant numbers 11135003, 11275246, and 11475238 (TL), and by the DOE grant DE-FG02-13ER42020 (DVN). 

\end{acknowledgments}

\end{document}